\documentstyle[preprint,revtex]{aps}
\begin{document}
\draft
\def\kjm{K_{j\alpha}}
\def\kjp{K_{jA}}
\def\kmi{K_{\alpha i}}
\def\kpi{K_{Ai}}
\def\kpm{K_{A\alpha}}
\def\wpm{W^{\uparrow}_{A\alpha}}
\def\wmp{W^{\downarrow}_{\alpha A}}

\def\G{{\cal G}}
\def\S{\left[ \sum_{n=0}^{\infty} (\kpm \wmp)^n \right]}
\begin{title}
The Quantum Propagator for a Nonrelativistic Particle \\
in the Vicinity of a Time Machine
\end{title}
\author{Dalia S. Goldwirth}
\begin{instit}
Center for Astrophysics,  60 Garden Street, Cambridge, MA 02138
\end{instit}
\author{Malcolm J. Perry}
\begin{instit}
DAMTP, University of Cambridge, Silver Street,
Cambridge, CB3 9EW, England
\end{instit}
\author{Tsvi Piran}
\begin{instit}
Center for Astrophysics,  60 Garden Street, Cambridge, MA 02138
\end{instit}
\author{Kip S. Thorne}
\begin{instit}
Theoretical Astrophysics, California Institute of Technology,
Pasadena, CA 91125
\end{instit}
\begin{abstract}
We study the propagator of a
non-relativistic, non-interacting particle in any non-relativistic
``time-machine'' spacetime of the type shown in Fig.~1: an external,
flat spacetime in which two spatial regions, $V_-$ at time $t_-$ and $V_+$
at time $t_+$, are connected by two temporal wormholes, one leading from
the past side of $V_-$ to t
the future side of $V_+$ and the other from
the past side of $V_+$ to the future side of $V_-$.  We express the
propagator explicitly in terms of those for ordinary, flat spacetime
and for the two wormholes; and from that expression we show that the
propagator
satisfies completeness and unitarity
in the initial and final ``chronal regions''
(regions without closed timelike curves) and its propagation from the
initial region to the final region is unitary.
However, within the time machine it satisfies neither
completeness nor unitarity.  We also give an alternative proof of
initial-region-to-final-region unitarity based on a conserved current
and Gauss's theorem.  This proof
can be carried over without change to most any non-relativistic
time-machine spacetime;  it is the non-relativistic version of a
theorem by Friedman, Papastamatiou and Simon, which says that for a free
scalar field, quantum mechanical unitarity follows from the fact that
the classical evolution
preserves the Klein-Gordon inner product.

\end{abstract}

\narrowtext

\section{introduction}

Spacetime and the phenomenon of gravitation are described very well at a
classical level by the theory of General Relativity.  Locally, spacetime is
isomorphic to Minkowski space and there is a well defined light cone and
microscopic causality.  Globally, however, things may be quite different.
There is nothing in the laws of classical general relativity that prevents
spacetimes from having closed causal (timelike or null) curves, that is
future directed curves through a point $p$ such that if one travels along
them always towards the local future, one returns to the same
spacetime point.  It is
easy to find examples of spacetimes in which closed timelike curves have
always existed \cite{r1}.  None of these examples, generally referred to as
eternal time machines, look very much like our Universe. In each of these
cases, it is not possible to pose the Cauchy problem in the usual
manner, at an arbitrary
``initial moment of time,'' for matter fields
propagating in these spacetimes \cite{r5}, and one can therefore expect
that these spacetimes are rather pathological.

Another type of causality violation is one in which closed timelike
curves develop during the evolution of a spacetime from some
reasonable initial conditions. An example of such behavior is found in
the Kerr solution which is believed to be the endpoint of
gravitational collapse with rotation. The region in which causality
violation occurs is close to the singularity and interior to the inner
horizon.  It might be the case that such behavior is
generic under
certain circumstances, as Tipler \cite{r7} has shown that if matter
obeys the weak energy condition, and closed timelike curves develop to
the future of some Cauchy surface, then the spacetime must be
geodesically incomplete.  We conjecture that if the weak energy
condition is satisfied in an asymptotically flat spacetime,
the closed timelike curves will also only occur
in the interiors of horizons and the physics exterior to any horizon
will always be unaffected by the paradoxes and difficulties
associated with closed timelike curves. Hawking \cite{r9} has proposed
the Chronology Protection Conjecture, presently still unproven, which
would prevent closed timelike curves
under a wide range of circumstances.

Systems that obey the weak energy condition classically, for example
a free scalar field, do not necessarily obey it after quantization
\cite{r10,r11}. Under these circumstances, it may be possible
to create a region of spacetime that includes closed timelike curves
(a {\it dischronal region})
without the occurrence of spacetime singularities other than those
associated with the chronology horizon.  Similarly, since the laws of
physics are time reversal invariant, we expect that dischronal regions could
disappear.  Any spacetime in which a dischronal region is preceded and
followed by {\it chronal regions} (regions without closed timelike
curves) will be referred to in this paper as a ``time machine."
Morris, Thorne and Yurtsever \cite{r12} have shown that one way such
spacetimes can arise is from the relative motion of the mouths of
a spatial wormhole that initially connects
two spacelike separated regions in Minkowski space.

There are a number of undesirable, apparent paradoxes
that arise in such a spacetime.  Recently several authors
\cite{r13,r14} have discussed the resolutions of some of
these paradoxes within the
realm of classical physics.  Despite these resolutions,
the Cauchy problem fails to be well-posed for classical interacting systems
(e.g.\ ``billiard balls'') in the presence of time machines
\cite{r13}: For the classical initial value problem, there can exist an
infinite number of consistent (i.e. nonparadoxical) classical
evolutions.  In other words, although the paradoxes can be avoided,
predictability is still violated, classically.

Not so quantum mechanically.  Quantum mechanics restores predictability
in the usual, probabilistic sense \cite{ra,r23,rb}: each of the
allowed classical evolutions acquires, in the WKB approximation, a
finite, predictable probability of being followed.  However, predictability is
restored only at a price: in the presence of a time machine, quantum
mechanics does {\it not} retain all of the ``nice'' features that we
normally are accustomed to.  For example, the propagator which takes an
evolving system from initial conditions before the dischronal region to
a final state afterward is {\it not}, in general, unitary
\cite{r23,rb,r24,r22}.  On the other hand, when the system being
evolved is non-interacting (free), the evolution {\it is} unitary, at least for
the those examples that have been studied:  Friedman, Papastamatiou,
and Simon \cite{r23} have shown that
a relativistic, free scalar field evolves unitarily
in {\it any} time machine
spacetime that is initially static and finally
static; and Politzer \cite{r22}
has proved unitarity for
a non-relativistic, free particle in a particular time machine:
flat spacetime with identification of
identical, finite sized regions at different times.

The generality of unitarity for a free scalar field suggests that,
similarly, the non-relativistic free particle should evolve unitarily in most
any time-machine spacetime.  That it does, indeed, do so we shall
demonstrate in this paper.

Among all standard formulations of quantum mechanics, the conceptually simplest
and most familiar is the Hamiltonian formulation in the
Schr\"odinger picture; there
the state at time $t$, $|\psi
(t)\rangle $, is determined by evolving
an initial state $|\psi (0)\rangle$ with the Hamiltonian operator
$H(t)$. We cannot use this Hamiltonian
formulation in the presence of a time machine (or more generally in
any non-globally-hyperbolic spacetime), because such a spacetime cannot
be globally foliated by spacelike surfaces of constant ``time'' $t$, and
correspondingly the standard
notion of ``state at time $t$,'' $|\psi (t)\rangle$, does not
exist; see, e.g.,\ \cite{rb} and references therein.

The only standard formulation of quantum mechanics that seems to survive in a
non-globally-hyperbolic spacetime is Feynman's path-integral
formulation \cite{r23,rb}. The
path-integral formulation can be derived from the Hamiltonian formulation
and conversely,
for
certain classes of Hamiltonian, provided that the Hamiltonian formulation
exists \cite{r15}.  However, in the absence of a Hamiltonian
formulation, the path
integral is the only foundational tool available for quantum theory.

In this paper we shall
study a free (non-interacting), non-relativistic particle.  We begin by
recalling a few key features of such a particle's
quantum mechanical description in a
globally hyperbolic spacetime (no time machine).  Suppose that
at a time $t_i$, the particle is
in an eigenstate of position at $x_i$, which we denote $|i, t_i
\rangle $. The propagator then is the amplitude ${\cal G}_{ji}$ given by
\begin{equation}
{\cal G}_{ji} = \langle j,t_j | i, t_i \rangle= \sum \exp [{ i S_{ji} /
\hbar }]
\label{Gij}\end{equation}
where the summation is over all paths from $(x_i, t_i)$ to  $(x_j,t_j)$
and $S_{ji}$ is the classical action evaluated along the path in question.

In a globally hyperbolic spacetime,
the propagator
obeys the group properties of completeness
\begin{equation}
{\cal G}_{ji} = \sum_k {\cal G}_{jk} {\cal G}_{ki}  \ \ \ \ \ t_j \ge t_k
\ge t_i
\label{comp}\end{equation}
and unitarity
\begin{equation}
{ \sum_k {\cal G}_{ki}^*{\cal G}_{kj} =
\cases { \delta_{ij}, & if \ \ \ $t_i=t_j < t_k$ \cr
{\cal G}^*_{ji}, & if \ \ \      $t_i < t_j < t_k$ \cr
{\cal G}_{ij}, & if \ \ \     $ t_j < t_i < t_k$ \ . \cr}}
\label{unit}\end{equation}
Completeness asserts that if one examines ${\cal G}_{ik}$, then the
particle will have been at some position at any intermediate time
$t_j$.  Unitarity is the statement that it is possible to reverse the
time evolution of a system so as to reconstruct an earlier state of
the system given the state at a later instant of time. In a globally
hyperbolic spacetime, unitarity can
be viewed as equivalent to conservation of probability.  It
should be noted that completeness and unitarity ensure that the time
evolution of a system is described by elements of a group, since the
additional axiom of associativity is clearly satisfied as a
consequence of (\ref{comp}).  In the Hamiltonian formulation,
completeness and unitarity are trivially guaranteed by Hermiticity of
the Hamiltonian, as
$H$ is the generator of the Lie algebra associated with the group of time
evolution.

As an explicit example consider the propagator $K_{ji}$ of a free
non-relativistic particle of mass $m$ propagating in the standard
flat spacetime of non-relativistic physics \cite{r15}:
\begin{equation}
K_{ji} = \cases { \left ( { m \over 2 \pi i \hbar (t_j-t_i) }
\right )^{{3\over 2}}
\exp \left ( { i m (x_j - x_i)^2 \over 2 \hbar (t_j-t_i) } \right ),
& if $t_j > t_i$ \cr
&     \cr
0, & if  $t_j < t_i$ \ . \cr}
\label{free}\end{equation}
$K_{ji}$ vanishes if $t_j < t_i$ since
the non-relativistic particle propagates
only to the future. Since we are dealing with a non-relativistic
propagator, the light cone is the line $t= \hbox{\rm const}$; i.e., a particle
located at $(x_i,t_i)$ can propagate to any point in which $t>t_i$.
Clearly $K_{ji}$ obeys (2) and (3), and can be derived by either
Hamiltonian methods or by path integrals.

In a time-machine spacetime, if
the evolution from time $t_i$ in the initial chronal
region, through
the dischronal region, to time $t_j$ in the final chronal region is
unitary, then
the standard notion of quantum mechanical state and the Hamiltonian
formulation of quantum mechanics exist in both chronal
regions
but not in the dischronal region \cite{r23,rb}, and the unitary
propagator ${\cal G}_{ji}$ relates the initial and final states in the
usual way [Eq.~(\ref{Gij})].  If the propagator is not
unitary, then its use and the
associated formulation of quantum mechanics might be slightly different from
what one is accustomed to in globally hyperbolic situations \cite{r23,rb},
but this will not concern us here.

Politzer \cite{r22} has shown, for a particular time machine,
that the non-relativistic,
free-particle propagator is unitary.
In this paper we generalize his result to the broader class of
time-machine spacetimes depicted in Fig.~1.
Each such spacetime consists of a standard, non-relativistic, flat
exterior region
plus two temporal wormholes.  The wormholes
connect two different, arbitrarily shaped spatial regions in the
exterior spacetime, $V_-$ at time
$t_-$ and $V_+$ at time $t_+ > t_-$.  The
upper wormhole in the figure connects the bottom (past) face of $V_+$
to the top (future) face of $V_-$; i.e.\ it is a wormhole in which, by
traveling forward in local time, one travels backward in external time
from $t_+$ to $t_-$.  The lower wormhole connects the bottom
(past) face of $V_-$ to the top (future) face of $V_+$, so that by
traveling forward in local time through it, one jumps forward in
external time from $t_-$ to $t_+$.  The shapes of the wormholes are
arbitrary and the space inside them can be curved and time-evolving.  They
might have perfectly reflecting walls, or points on
their ``walls'' might be identified in such a way that the
wormholes are spatially closed with no real walls at all.  Whatever may
be their form, because the spacetime inside them is assumed to be
nonsingular and foliable by a family
of spacelike hypersurfaces,
the path integral will produce a unitary
propagator that takes the
particle forward in local time from
the beginning of each wormhole to its end.

We introduce the notation
that Greek indices $\alpha,\beta...$ denote points in $V_-$ (at $t_-$) and
capital Latin indices $A,B,...$ denote points in $V_+$ (at $t_+$),
and we denote the unitary propagator from $(x_B,t_+)$
through the upper wormhole to $(x_\beta,t_-)$ by $W^{\downarrow}_{\beta
B}$, and the unitary propagator from $(x_\alpha,t_-)$ through the lower
wormhole to $(x_A,t_+)$ by $W^{\uparrow}_{A\alpha}$.  The arrows on
these propagators indicate the direction of propagation relative
to exterior time. Politzer's time machine
is obtained by choosing $V_-$ and $V_+$ to be
identical in size and shape, and choosing the wormholes to be
vanishingly short so the future side of $V_-$ is identified with
the past side of $V_+$ and conversely.  In Politzer's
time machine the wormhole propagators
degenerate to the identity function,
$W^{\downarrow}_{\alpha A}= W^{\uparrow}_{A \alpha} = \delta_{A \alpha}$.

In Sec.~II of this paper, we construct the propagator ${\cal G}_{ji}$ for
any wormhole of the type shown in Fig.~1, expressing it in terms of the
ordinary flat-spacetime propagator $K_{ji}$ and the interior propagators
$W^{\downarrow}_{\alpha A}$ and $W^{\uparrow}_{A \alpha}$, and we
use this expression for ${\cal G}_{ji}$ to prove that it is
complete and unitary in the initial and final chronal regions, and propagates
unitarily from the initial to the final region. Although
the calculation presented uses the flat spacetime propagator, it is
straightforward to see that $K_{ji}$ can be replaced by any unitary
propagator and our results still remain valid. Therefore we can strengthen
our conclusions to include a much wider range of spacetimes; namely those
where, both prior and subsequent to the dischronal region, the non-relativistic
propagator is unitary, as would arise under almost any circumstance.

In Sec.~III we show that, for the spacetime of Fig.~1, ${\cal G}_{ji}$ obeys
the
Schr\"odinger equation everywhere (in the dischronal region and
the wormholes as well as the chronal regions), and we use that fact plus
Gauss's theorem to prove
unitarity.  This second demonstration of unitarity has the virtue that
it generalizes without change to most any time machine.
In Sec.~IV we present concluding remarks.

\section{Explicit Expression for the Propagator, and its Properties}

Had there been no wormholes, then the propagator ${\cal G}_{ji}$
in the time-machine
spacetime would simply become $K_{ji}$ given by Eq.~(\ref{free}).
However, it is not too hard to evaluate the path integral (\ref{Gij})
explicitly in the time-machine spacetime which we are considering, because
it is simple to find all possible paths by which a particle can propagate
from $(x_i,t_i)$ to $(x_j,t_j)$. We will explicitly calculate ${\cal
G}_{ji}$ for the case that $t_i < t_- < t_+ < t_j$. The possible paths are
then labeled by the number of times $n$ that the particle traverses a
wormhole, and the contribution to ${\cal G}_{ji}$ from all paths with fixed
$n$ is ${\cal G}_{ji}^{(n)}$.  For $n=0$ we have:
\begin{eqnarray}
{\cal G}^{(0)}_{ji}= K_{ji} - \int_{V_+} d^3 x_A K_{jA} K_{Ai} -
\int_{V_-} d^3 x_\alpha K_{j\alpha} K_{\alpha i}
+  \int_{V_-} d^3 x_\alpha \int_{V_+} d^3 x_A K_{jA} K_{A\alpha} K_{\alpha i}
\nonumber\\
\equiv K_{ji} - K_{jA}K_{Ai}  - K_{j\alpha }K_{\alpha i} +
K_{jA} K_{A\alpha} K_{\alpha i}\  ,
\label{G0}
\end{eqnarray}
where $K_{ji}$ is the ordinary propagator (\ref{free})
in the flat spacetime
and where in the second equality we adopt the
notation, like the summation convention, that repeated {\it adjacent}
indices are integrated over (thus the indices behave like matrix
indices).  The first term comes from all paths that go from
$(x_i,t_i)$ to $(x_j, t_j)$. The second and third terms, which
represent the contribution of all paths from $(x_i,t_i)$ to $(x_j,
t_j)$ via $V_+$, and all paths from $(x_i,t_i)$ to $(x_j, t_j)$ via
$V_-$ respectively, must be subtracted off, since any particle that
arrives at $(x_A,t_+)$ will then enter a wormhole instead of traveling
directly on to $(x_j,t_j)$, and similarly for particles arriving at
$(x_\alpha,t_-)$.  In these subtractions we have double counted the
paths that in ordinary space would have gone from $(x_i,t_i)$ via
$V_-$ to $V_+$ and then to $(x_j,t_j)$, so those are added in the last
term.

For $n=1$ the calculation can be done in much the same way:
\begin{eqnarray}
{\cal G}^{(1)}_{ji}=
K_{j\alpha} W^{\downarrow}_{\alpha A } K_{Ai}  -  K_{jA}  K_{A\alpha}
W^{\downarrow}_{\alpha B} K_{Bi}
- K_{j\alpha}  W^{\downarrow}_{\alpha A} K_{A\beta} K_{\beta i} \nonumber \\
+K_{jA} K_{A \beta} W^{\downarrow}_{\beta B} K_{B \alpha} K_{\alpha i}
+ K_{jA} W^{\uparrow}_{A\alpha } K_{\alpha i}\ .
\label{G1}\end{eqnarray}
The first term in
(\ref{G1}) is the contribution from a particle traveling once through the upper
wormhole.  The second, third, and fourth are the same types of
corrections as we met in Eq.~(\ref{G0}). The last
term is the contribution of paths that reach $(x_\alpha,t_-)$,
then travel through the lower wormhole to $(x_A,t_+)$ and from there
to $(x_j,t_j)$.

Similarly one can construct the general ${\cal G}^{(n)}_{ji}$ for paths
with $n$ wormhole traversals.
Unlike the previous two cases, for $n
\ge 2$ we have contributions only from paths that begin by reaching
$(x_A,t_+)$:
\begin{eqnarray}
{\cal G}^{(n)}_{ji}= K_{j\alpha} W^{\downarrow}_{\alpha A} \underbrace{
K_{A\beta}
W^{\downarrow}_{\beta B} K_{B\gamma} W^{\downarrow}_{\gamma C}....
K_{C \delta } W^{\downarrow}_{\delta D}}_{{\rm (n-1) \ times}} K_{D i}
\nonumber\\
- K_{jA} \underbrace{K_{A \alpha } W^{\downarrow}_{\alpha B} K_{B \beta}
W^{\downarrow}_{\beta C}....
K_{C \gamma} W^{\downarrow}_{\gamma D}}_{{\rm n \ times}} K_{D i} \nonumber\\
+ K_{j A}  \underbrace{ K_{A \alpha } W^{\downarrow}_{\alpha B}
K_{B \beta } W^{\downarrow}_{\beta C}....
K_{C \gamma } W^{\downarrow}_{\gamma D}}_{{\rm n \ times}}K_{D \delta}
K_{\delta i}
\nonumber\\
- K_{j \alpha } \underbrace{W^{\downarrow}_{\alpha B} K_{B \beta}
W^{\downarrow}_{\beta C}
K_{C \gamma}....
W^{\downarrow}_{\gamma D} K_{D \delta}}_{{\rm n \ times}} K_{\delta i}
\ \ \ .
\label{Gn} \end{eqnarray}

The complete propagator can now be evaluated in terms of $K_{ji}$ by
summing over all ${\cal G}^{(n)}$:
\def\S{[ \sum_{n=0}^{\infty} (K W^{\downarrow})^n ]}
\begin{eqnarray}
{\cal G}_{ji} =  \sum_{n=0}^\infty {\cal G}^{(n)}_{ji} && = K_{ji}
- K_{j\alpha} \left \{ \delta_{\alpha \beta}
+ W^{\downarrow}_{\alpha A} \left [ \sum_{n=0}^\infty (K W^{\downarrow})^n
\right ]_{AB}
K_{A \beta}  \right \} K_{\beta i}
- K_{j A} \left [ \sum_{n=0}^\infty (KW^{\downarrow} )^n \right ]_{AB}
K_{Bi}  \nonumber\\
&& + K_{j A}  \left \{ W^{\uparrow}_{A \alpha}  +
\left [ \sum_{n=0}^{\infty} (K W^{\downarrow})^n \right ]_{BC}
K_{C \alpha} \right \}  K_{\alpha i}
+ K_{j \alpha} W^{\downarrow}_{\alpha A} \left [ \sum_{n=0}^{\infty}
( K W^{\downarrow})^n \right ]_{AB} K_{Bi}
\ .
\label{Gijfa}\end{eqnarray}
The sum $[ \sum_{n=0}^{\infty} ( K W^{\downarrow})^n ]_{AB}$ can be
expressed formally as:
\begin{equation}
\left [ \sum_{n=0}^{\infty} ( K W^{\downarrow})^n \right ]_{AB} =
\left [{ 1 \over \delta - K W^{\downarrow}} \right ]_{AB}
\end{equation}
Collecting terms and using this definition  we obtain:
\begin{equation}
{\cal G}_{ji} =  K_{ji}  +
(K_{j \alpha} W^{\downarrow}_{\alpha A} - K_{jA} ) \left [{ 1 \over \delta - K
W^{\downarrow}} \right ]_{AB}
(K_{Bi} - K_{B\beta} K_{\beta i} )
+ (K_{j A} W^{\uparrow}_{A \alpha} - K_{j\alpha}) K_{\alpha i}\ .
\label{Gijf}\end{equation}

The propagator ${\cal G}_{ji}$ was derived assuming that $t_i < t_-$ and that
$t_j > t_+$. However, (\ref{Gijf}) holds for all values of $t_i$ and
$t_j$, as
can readily be seen by considering all possible time orderings of $t_i$
and $t_j$ relative to $t_-$ and $t_+$ and by noting that $K_{ba}=0$ if
$t_b < t_a$.  In other words, {\it the
path-integral-defined propagator ${\cal G}_{ji} = \sum
\exp[iS_{ji}/\hbar]$ can be expressed in the form (\ref{Gijf}) for all
pairs of points $i,j$ that reside outside the wormholes}.

Using the same method, we can express the time reversed operator,
${\cal G}^{(R)}_{ij}$, which is the propagator to go backwards in time
from the point $j$ to $i$, in terms of $K^{(R)}_{ij}$,
$W^{\downarrow (R)}_{A \alpha}$
and $W^{\uparrow (R)}_{\alpha A}$. By definition, ${\cal G}^{(R)}_{ij}$
is the sum of $\exp(iS_{ij}/\hbar )$ over all paths that start at the point $j$
and
travel backwards in time to the point $i$, and similarly for
$K^{(R)}_{ij}$, $W^{\downarrow (R)}_{A \alpha}$
and $W^{\uparrow (R)}_{\alpha A}$.  The standard argument, valid
whenever the relevant region of spacetime is foliable by spacelike
hypersurfaces (so no issues of closed timelike curves arise), shows that
$K^{(R)}_{ij} = K^*_{ji}$
(i.e.\ time-reversed propagation is produced by the  Hermitian conjugate
of $K_{ji}$), and similarly
$W^{\downarrow (R)}_{A \alpha}=W^{\downarrow *}_{\alpha A}$ and
$W^{\uparrow (R)}_{\alpha A} = W^{\uparrow *}_{ A\alpha}$. These
relations for the propagators are referred to as ``Hermiticity.''  Our
computation reveals that ${\cal G}^{(R)}_{ij}$ is given in terms
of $K^{(R)}_{ij}$, $W^{\downarrow (R)}_{A \alpha}$
and $W^{\uparrow (R)}_{\alpha A}$ by an expression identical in form to
Eq.~(\ref{Gijfa}); and this, together with Hermiticity of $K_{ji}$,
$W^{\downarrow}_{\alpha A}$ and $W^{\uparrow}_{ A\alpha}$, implies that
${\cal G}^{(R)}_{ij} ={\cal G}^*_{ji} $.  Thus, ${\cal G}$ is Hermitian.

To examine the completeness properties of ${\cal G}_{ji}$ we need to
evaluate $\sum_k {\cal G}_{jk} {\cal G}_{ki}$ for the various cases of
$t_j$, $t_k$ and $t_i$ greater or less than $t_-$ and $t_+$ and subject to
$t_i < t_k < t_j$. We use the completeness properties (\ref{comp}) of $K_{ji}$,
together with $\sum_k K_{Ak} K_{kA} = 0$. The latter follows
from the fact that either $K_{Ak}$ or $K_{kA}$ vanishes depending on
whether $t_k$ is greater than or less than $t_+$.  If $t_k>t_+$ or $t_k <
t_-$, completeness of ${\cal G}_{ji}$ follows directly from the
completeness of $K_{ji}$.  Hence, ${\cal G}_{ji}$ obeys the completeness
condition if the intermediate surface $(t=t_k)$ is chosen to be either to
the past or the future of the time machine. If however $t_- < t_k < t_+$
then
\begin{equation}
\sum_k {\cal G}_{jk} {\cal G}_{ki}\ne {\cal G}_{ji}
\end{equation}
and completeness fails to be satisfied. This is because the particle can
cross such an intermediate surface exterior to the wormholes any number of
times, including zero.  This violation of completeness seems
harmless as it happens only
while the time machine is operating and completeness is recovered after the
time machine has ceased to exist.

Because ${\cal G}_{ji} = K_{ji}$ for $(x_i,t_i)$ and $(x_j,t_j)$ both in
the initial chronal region or both in the final chronal region, ${\cal
G}_{ji}$ is unitary in each of these regions.
To check unitarity for $(x_i,t_i)$ in the initial chronal region and
$(x_j,t_j)$ in the final chronal region, i.e., for propagation through
the time machine,
we substitute the full expression for ${\cal G}^*_{kj}$
and ${\cal G}_{ki}$, given by  (\ref{Gijf}) into the unitarity sum
$\sum_k {\cal G}^*_{kj} {\cal G}_{ki}$ to obtain:
\begin{eqnarray}
\sum_k {\cal G}^*_{kj} {\cal G}_{ki} =
\left [ K^*_{kj}  +
(K^*_{k \alpha} W^{*\downarrow}_{\alpha A} - K^*_{kA} )
\left ({ 1 \over \delta - K^* W^{*\downarrow}} \right )_{AB}
(K^*_{Bj} - K^*_{B\beta} K^*_{\beta j} ) \right.
\nonumber\\
\left . +
(K^*_{k A} W^{*\uparrow}_{A \alpha} - K^*_{k\alpha}) K^*_{\alpha j} \right ]
\nonumber \\
\left [K_{ki}  +
(K_{k \gamma} W^{\downarrow}_{\gamma C} - K_{kC} )
\left ({ 1 \over \delta - K W^{\downarrow}}\right )_{CD}
(K_{Di} - K_{D\delta} K_{\delta i} )  \right .
\nonumber \\
\left .
+ (K_{k C} W^{\uparrow}_{C \gamma} - K_{k \gamma}) K_{\gamma i} \right ]
\ \ \ .
\end{eqnarray}
The summation over $k$ appears between terms such as $K^*_{kj} K_{k \alpha}$,
for which we use the fact that $K$ is a flat space propagator and is
therefor itself unitary, to obtain
$\sum_k K^*_{kj} K_{k \alpha} = K_{j \alpha}$. Using this, the unitarity
of $W^{\uparrow}_{A\alpha}$ and $W^{\downarrow}_{\alpha A}$, and
the identity
\begin{equation}
(\delta_{AB} - K_{A \alpha} W^{\downarrow}_{\alpha B} )
\left ({ 1 \over \delta - K W^{\downarrow}} \right )_{BC} = \delta_{A C}
\ \ \ ,
\end{equation}
we discover that
the unitarity condition is satisfied for arbitrary $W^{\downarrow}$
and $W^{\uparrow}$.

Although ${\cal G}_{ji}$ propagates unitarily from the initial
chronal region to the final chronal region, its propagation within the
time machine (i.e.\ within the dischronal region) is not unitary,
as one can check most easily by choosing
$t_- < t_i = t_j < t_k < t_+$.

\section{Unitarity for a Broad Class of Time Machines}

We now present an
alternative proof of unitarity for propagation from the
initial chronal region to the final chronal region.
This proof (which is a generalization of one given in the unpublished
Ref.\ \cite{ra}) has the virtue that it is valid for any time machine whose
spacetime everywhere is locally (not globally) foliable by spacelike
hypersurfaces with proper time separations that are independent of
spatial location.  (This is the typical form of non-relativistic
spacetimes.)
If the local spatial coordinates are carried
perpendicularly from one hypersurface to the next, then the spacetime
metric takes the form
\begin{equation}
ds^2 = -c^2 d\tau^2 + g_{pq}(x,\tau)dx^pdx^q\ .
\label{metric}
\end{equation}
Here $c$ is the speed of light (which is regarded as arbitrarily large
since we are in the non-relativistic limit).  The time-machine spacetime
of Fig.~1 has this metric, with $g_{pq} = \delta_{pq}$ and $\tau = t$
in the flat exterior, but not inside the wormholes.  Note that, because $\tau$
everywhere increases toward the local future, it is not possible to
cover the dischronal region by a single coordinate patch of this sort;
several are required.  We assume (for conceptual simplicity) that, as in
Fig.~1, so also for our more general time machine,
in the distant-past portion of the initial chronal
region, space is flat; and similarly for the distant future of the
final chronal region.

In ordinary flat space,
the path-integral-defined propagator $K_{ai}$ satisfies the Schr\"odinger
equation in its final point $a$,
\begin{equation}
-{\hbar\over i}{\partial K_{ai}\over\partial t_a} = -
{\hbar^2\over2m}{\nabla_a^2 K_{ai}}\ .
\label{schrodinger_flat}
\end{equation}
The same infinitesimal-propagation
argument that is used to derive this equation in flat space (Sec.~4-1 of
Feynman and Hibbs \cite{r15}) can be used in spacetimes with the metric
(\ref{metric}) to derive the corresponding Schr\"odinger equation
\begin{equation}
-{\hbar\over i}{1\over g^{1/4}}{\partial g^{1/4}{\cal G}_{ai}\over\partial
\tau_a} = - {\hbar^2\over2m}{\nabla_a^2 {\cal G}_{ai}}\ .
\label{schrodinger_curved}
\end{equation}
Here $g \equiv \hbox{\rm det} ||g_{pq}||$ is the determinant whose
square root governs spatial volume elements, $\nabla_a^2$ is the
covariant, spatial Laplacian at the final point $a$, and there is no sum
over the point $a$. It should be emphasized that this equation was
derived by a local analysis, and it is not necessarily the case
that the right-hand side of (\ref{schrodinger_curved})
can be identified with a well-defined
global Hamiltonian for the system.
The Schr\"odinger equation (\ref{schrodinger_curved})
is valid inside the time machine as well as outside; closed
timelike curves do not affect its path-integral-based derivation.
To see that this is indeed the case, recall that
Eq.~(\ref{schrodinger_curved}) is derived by just varying the final
endpoint of the path integral. Therefore all that is required is that
the spacetime be regular in an infinitesimal neighbourhood of the point
in question.

If the particle were not free, its self-interactions
in the dischronal region
(e.g., billiard-ball collisions \cite{r13}) would produce contributions to the
action that invalidate the derivation of the Schr\"odinger equation
\cite{ra},
and presumably thereby
would invalidate the following proof of unitarity.

 From the Schr\"odinger equation (\ref{schrodinger_curved}), one can
easily derive the following differential conservation law, which is intimately
related to the conservation of the probability current:
\begin{equation}
\nabla_a \cdot \left [ \frac{i\hbar}{2m} ( {\cal G}_{ai} \nabla_a
{\cal
G}^*_{aj} - {\cal G}^*_{aj}\nabla_a {\cal G}_{ai} ) \right ]
+ \frac{1}{g^{1/2}}
\frac{\partial (g^{1/2} {\cal G}^*_{aj} {\cal G}_{ai} )}{\partial \tau_a}
=0\ .
\label{diff_cons}
\end{equation}
Here $\nabla_a$ is the covariant spatial gradient in the metric $g_{pq}$
and there is no sum over the point $a$.
We choose the points $i$ and $j$ to lie in the initial, flat-space
chronal region at times $t_i$ and $t_j$; and we let $t_m$ be a
time to the future of $t_i$ and $t_j$ but still in the initial, flat
chronal region,
and $t_k$ be a time in the final, flat chronal region.
We then construct the volume integral of the conservation law
(\ref{diff_cons}) over the spacetime region between $t_m$ and $t_k$,
apply Gauss's theorem to convert the spacetime volume integral into
a surface integral, and thereby obtain an integral conservation law.
(Recall that Gauss's theorem is valid independently of
the topology of the spacetime; it only requires orientability
and the existence of a metric).
The surface integral is over all
the boundaries of the integration 4-volume, and these include,
in addition to the initial $t_m$ and final $t_k$ surfaces, also any
walls such as those that might bound the temporal wormholes of Fig.~1.
If (as we assume) all such walls are perfectly reflecting, then they
give zero contribution to the surface integral, so the only surviving
contributions are from $t_m$ and $t_k$, and the resulting integral
conservation law takes the form
\begin{equation}
\sum_k {\cal G}^*_{kj} {\cal G}_{ki} = \sum_m {\cal G}^*_{mj} {\cal
G}_{mi} \ .
\label{int_cons}
\end{equation}
Since the times $t_m$, $t_i$, and $t_j$ are all in the initial, flat,
chronal region, the propagators on the right-hand side of this equation
are precisely the flat propagators $K^*_{mj}$ and $K_{mi}$; and their
unitarity  brings Eq.~(\ref{int_cons}) into precisely the same
expression as Eq.~(\ref{unit}).
Since the sum (integral) over $k$ in the resulting Eq.~(\ref{unit}) is
performed at time $t_k$ to the future of the time
machine, and the times $t_i$ and $t_j$ are to its past, this equation
states that propagation through the time machine is unitary.

However, we must make a cautionary remark about the above theorem.
As Tipler and Hawking have shown \cite{r7,r9}, whenever a
dischronal region arises in an asymptotically flat spacetime
generically, it
must be accompanied by
some form of spacetime singularity (though in some cases
the singularity can be exceedingly mild and irrelevant for
physics \cite{r12,r9}).  If in the derivation
of equation (\ref{int_cons}) the region we integrate over
contains a singularity, then the above derivation, strictly speaking,
is invalid.
The difficulty can, under a wide range of circumstances,
be fixed up by the following method. One simply needs to excise
the singular region and construct a boundary around
where the singular region was removed. On that boundary, one again sets
reflecting
boundary conditions, and the theorem we proved above will remain true.
What then is the physical significance of such a procedure?
In a certain sense it follows naturally from the physical motivation
for regarding the path integral as fundamental.
We are simply asserting that paths cannot begin or terminate on
the singularity. In fact, we have already used this condition
implicitly in the analysis of the
previous section. The surfaces of constant $t$ that contain
$V_-$ and $V_+$ (Fig.~1) possess \lq\lq mild'' singularities where the wormhole
mouths join onto the remainder of the flat space.  We dealt
with the problem there by simply requiring paths to be
continuous and to enter (leave)
the wormhole or to (have) avoid(ed) it. These conditions
are precisely the reflecting boundary conditions
required to make the theorem sketched above work.
In fact, any time machine resembling Fig.~1
will necessarily have singularities
of at least this type where the wormholes join the chronal region.
Whilst it is physically clear how to deal with such a mild type of
singularity, it may be that our method fails for more severe
types of singularity that one could imagine finding.

In essence, the analysis that we have given in this section
shows that unitarity is a consequence of the
local conservation law for probability current [Eq.~(\ref{diff_cons})].
This is a
non-relativistic variant of the theorem, by Friedman, Papastamatiou, and
Simon \cite{r23}, that for a free, relativistic scalar field, unitarity of
propagation through a
time machine follows from conservation of the Klein-Gordon inner
product (a theorem that is subject to cautionary remarks about singularities
similar to those stated above).

In our non-relativistic case
\cite{r22,ra}, as for a free, relativistic scalar field \cite{r23,r24},
self-interactions will break the conservation law and produce nonunitarity.

\section{Conclusions}

There are two obvious extensions that deserve examination.
The first is an extension of our explicit expressions and properties
for the propagator, in spacetimes similar to Fig.~1,
to the case of relativistic
free particles. In fact, it is more or less obvious that the techniques
of Sec.~II will carry over to the relativistic case with little change.
The only significant differences between
the relativistic and the non-relativistic cases are firstly, that
particles can only propagate to the future inside the light cone of
their starting point (and similarly for antiparticles traveling
backwards in time), and secondly, there may be numerous
particle-antiparticle creation or annihilation events which cause the
trajectory of the particle to zigzag. However, neither of these issues
will affect the basic structure of the calculations, and it would
appear that all one has to do is to replace $K_{ji}$ by the
appropriate relativistic propagator in flat spacetime in order to
obtain the corresponding relativistic results.

A second, vastly more complex problem, has to do with seeking a deeper
understanding of the loss of unitarity for interacting systems.
A violation of unitarity means that,
if one were to attempt
computing probabilities in both the initial and final chronal regions
using the Hamiltonian formulation's rules (i.e.\ the Copenhagen
interpretation), one would find that
probability is not conserved.  Hartle \cite{rb} and Friedman,
Papastamatiou, and Simon \cite{r23} respond to this by seeking from the
path-integral formalism and other considerations
an alternative way of computing probabilities.  From their
alternative way (which the fourth author of this paper finds compelling),
they conclude that, although one recovers the usual
Hamiltonian formulation of quantum mechanics from the path-integral
formulation to the future of the time machine, one cannot do so to the
past, even though the past region is chronal.

The first three authors
of this paper are inclined to believe a somewhat different
picture.
Since the spacetime has no boundaries (except at infinity) and is
non-singular, it is difficult to see why the standard probability current
should
fail to be conserved. The only physical explanation would be that
probability fails to be conserved because of the possibility of having
particles being eternally trapped inside the time machine (either to
the future or to the past).  Implicitly, we assumed that there are no
particles emerging from the time machine that were not put in
originally.  Therefore, the only potential source of difficulty would
be their becoming trapped forever in the future there. This is, by
the results of our construction, demonstrably not the case for free
particles. It appears that unitarity violation for interacting
particles in the spacetime of Fig.~1
arises due to self interaction just outside the mouths of the upper
wormhole,
and this causes a particle that emerges from the past mouth to
return to the future one. The overall result is that the particle becomes
trapped forever.

To summarize, we have seen in this paper that,
at least at the level of our
analysis, there is no contradiction between the postulates of quantum
mechanics and the possible existence of causality violation in
general relativity.  This is quite distinct from the situation in
classical mechanics, where the inability to solve uniquely an
apparently sensible initial value problem leads to a breakdown of
classical predictability. We believe that it is quite possible that a
suitable quantum mechanical treatment will resolve many of the
classical difficulties of causality violation under wider assumptions
than those treated here.

\acknowledgments

It is a pleasure to acknowledge S. Coleman, J. Friedman, J. Hartle, and
A. Strominger for enlightening conversations. We thank the Aspen Center for
Physics for hospitality while much of this research was
being done.  This research was
partially supported by a Center for Astrophysics fellowship (DSG),
by the  Royal Society and Trinity College (MJP), by the US-Israel
BSF (TP), and by Caltech's Feynman research fund (KST).

\newpage
\figure{
A non-relativistic time-machine spacetime and two paths in the
spacetime that contribute to
the propagator ${\cal G}_{ji}$ for a non-interacting, non-relativistic
particle.  {\it Left:} The external, non-relativistic, flat part of the
spacetime.  {\it Right:} The temporal wormholes which connect the
regions $V_-$ and $V_+$.
}

\end{document}